# The Auger Radioisotope Microscope: an instrument for characterization of Auger electron multiplicities and energy distributions

**Patrick R. Stollenwerk[1], Stephen H. Southworth[2], Francesco Granato[1], Amy Renne[1,3], Brahim Mustapha[1], Kevin G. Bailey[1], Peter Mueller[1], Jerry Nolen[1,3], Thomas P. O'Connor[1], Junqi Xie[1], Linda Young[2,4], and Matthew R. Dietrich[1]**

1 Physics Division, Argonne National Laboratory, Lemont, IL 60439, USA
2 Chemical Sciences and Engineering Division, Argonne National Laboratory, Lemont, IL 60439, USA
3 Department of Radiology, The University of Chicago, Chicago, IL 60637, USA
4 The James Franck Institute and Department of Physics, The University of Chicago, Chicago, IL 60637, USA

**E-mail:** nolen@anl.gov and pstollenwerk@anl.gov

## Abstract

We describe a new instrument, the Argonne Auger Radioisotope Microscope (ARM), capable of characterizing the Auger electron emission of radionuclides, including candidates relevant in nuclear medicine. Our approach relies on event-by-event ion-electron coincidence, time-of-flight, and spatial readout measurement to determine correlated electron multiplicity and energy distributions of Auger decays. We present a proof-of-principle measurement with the ARM using X-ray photoionization of stable krypton beyond the *K*-edge and identify a bifurcation in the electron multiplicity distribution depending on the emission of *K-LX* electrons. Extension of the ARM to the characterization of radioactive sources of Auger electron emissions is enabled by the combination of two recent developments: (1) cryogenic buffer gas beam technology to introduce Auger emitters into the detection region with well-defined initial conditions, and (2) large-area micro-channel plate detectors with multi-hit detection capabilities to simultaneously detect multiple electrons emitted in a single decay.

The ARM will generate new experimental data on Auger multiplicities that can be used to benchmark atomic relaxation and decay models. This data will provide insight into the low-energy regime of Auger electrons where intensity calculations are most challenging and experimental data is limited. In particular, accurate multiplicity data of the low-energy regime can be used to inform oncological dosimetry models, where electron energies less than 500 eV are known to be most effective in damaging DNA and cell membranes.

## 1. Introduction

Radioisotope Auger electron (AE) emitters have been identified as a promising source for targeted radionuclide therapy (TRT) *[1, 2]*. The goal of TRT is to deliver destructive radionuclides to cancerous cells without damaging nearby healthy cells. In a radionuclide AE emitter, a cascade of many electrons with energies much lower than typical beta, gamma, or alpha particles are shed from the decayed atom. This cascade of electronic emissions occurs with the formation of inner electronic shell vacancies which generally follow the absorption of an electron by the nucleus (i.e., electron capture) or are generated via the electromagnetic interaction of an excited nucleus and a core electron that results in the emission of a conversion electron (i.e., internal conversion). The low-energy, high-multiplicity nature of AEs means that much of the deposited energy is short-range (of order nm) compared to the cell size and has characteristically high overall linear energy transfer. Thus, AE emitters have the potential to satisfy the goals of TRT if they can be successfully delivered to sensitive targets of cancerous cells *[3, 4, 5, 6, 7]* such as the DNA or cell membrane.

Characterizing the details of the energy distribution and multiplicity of AEs following decay is critical for calculating the estimated damage, optimal dosage, and overall efficacy of a particular Auger electron emitting radionuclide candidate. It is the low-energy part of the spectrum (< 1 keV), where models are least reliable *[8, 9]*, which is most effective in causing localized damage. Electrons with an energy less than 500 eV are the most effective at inducing DNA double strand breaks *[10]* including substantial contribution from those below 30 eV *[11, 12, 13, 14, 15]*. As electron energies drop below 30 eV, their de Broglie wavelengths become comparable to interatomic and intermolecular dimensions in the condensed phase and consequently quantum effects can significantly modify their cross-sections and enhance the probability of inducing DNA double strand breaks *[14, 16, 17]*. Thus, a detailed understanding of the lowest energy region of the spectrum is key to predicting which radionuclides are the most effective at damaging cancerous tissue.

While a few Auger emitters have made it to Phase 1 clinical trials with effective therapeutic outcomes *[18, 19]*, total remissions have been rare. Uncertainty in translating pre-clinical results into dosimetry and toxicity for clinical trials has limited progress in applying more appropriate doses for better therapeutic outcomes *[20]*. Notably, Auger decays themselves are not well characterized. To date there have been very few direct experimental measurements of the electron multiplicities and low energy spectra of therapeutic Auger emitting isotopes. However, the highly localized dose deposited by Auger electrons calls for a well-understood emission spectrum to enable calculation of the radiation absorbed dose and subsequent therapeutic effectiveness assessments. There are a variety of Monte Carlo decay simulation techniques *[9, 10]*, the assumptions of which need to be validated or refined based on experimental data *[21]*. Typically, these codes rely on pre-calculated inputs from available databases, e.g., the Evaluated Atomic Data Library (EADL) *[22]*, and the absorbed dose in a sub-micron radius between dosimetry models can disagree by orders of magnitude due to differences in the input spectrum *[9]*. The actual verification of these Monte Carlo models and their underlying assumptions is very challenging given the inherent difficulty in measuring the yields of the lowest energy Auger electrons *[23]*.

Most experimental data on AEs have been recorded using either cylindrical mirror analyzers (CMAs) *[8, 24, 25]* or hemispherical *[26]* analyzers which can measure emission spectra energies over several orders of magnitude. In these instruments, high-resolution spectra are generated using slits which act as narrow band-pass filters that can be scanned across the spectrum to acquire relative intensities of emission lines. In this way, Alotiby et al. *[25]* compared conversion electron intensity to AE intensity to expose a 20% deviation of the expected *K*-Auger yield according to calculations assuming transition rates provided by the EADL. Such an approach, however, precludes measuring the multiplicity distribution on an event-by-event basis and requires dense, high-activity samples, typically on solid substrates, which make detection of the lowest energy electrons challenging to impossible. For this reason, most experiments only measure spectra well above 1 keV. State-of-the-art experiments have been able to measure spectra as low as 200 eV by using a monolayer film of $^{125}$I yet were still insensitive to the bulk of AEs which are at an even lower energy *[8]*. This approach is also not applicable for general purpose measurement of the low-energy spectrum as it requires the technically challenging task of preparing a monolayer of the radioactive sample *[27]*.

As experimental data from CMAs provides spectra with relative intensities, it fundamentally has limited insight into the distribution of events, e.g., the probability of any given number of emitted electrons per decay. This more complete data is important to determine the cytotoxicity of these isotopes and is currently provided only by theory. In a few cases of noble gas isotopes, multiplicity distribution data is available and has been compared to simulation *[28, 29]*. In this way discrepancies of the peak multiplicity between theory and experiment have been exposed in $^{131m}$Xe *[30]*. The theoretical multiplicity distribution consistently underestimated peak multiplicities and completely missed the highest ion charge states, possibly due to inaccuracies in the EADL. Depending on the level structure, an excited nuclei decay can result in a rapid cascade (<< 1 ns) of multiple internal conversions generating multiple vacancies. In the case of electron capture, the formation of a vacancy is accompanied by a decrease of $Z$ of the nucleus leaving the atom to be initially neutral. The change in $Z$ can also leave the nucleus in an excited state opening the possibility for generating more vacancies via internal conversion. To isolate the role of atomic relaxation models versus the initial vacancy and electronic structure assumptions in these limited noble gas cases, comparison of theory to experiment is usually done using photoionization data on stable species where experimental uncertainty in the distribution of initial atomic holes is much smaller. Though important for validating relaxation models, this benchmark multiplicity data is not directly comparable to the multiplicity expected from a spontaneous nuclear decay, motivating the need for direct benchmarking of radioactive species.

We describe a new apparatus, the Auger Radioisotope Microscope (ARM), capable of measuring the multiplicities and spectra of the low energy electrons of Auger electron emitting atoms decaying in-flight, under high vacuum. In contrast to current methods, the ARM is designed to measure the full Auger electron multiplicity and energy distribution event-by-event rather than in a scanning and relative/average mode. The combined, simultaneous energy and multiplicity measurements at the single decay level are accomplished through time-of-flight (TOF) ion-electron coincidence detection and enable the measurement of *detailed* multiplicities, i.e., the multiplicity data binned by electron energy. These detailed multiplicities therefore give us direct access to the distribution of electrons in biologically relevant energy regimes, namely those where cross sections are impacted by quantum effects ($\lesssim$ 30 eV), those which contribute mostly to local (~1-30 nm) biological effects ($\lesssim$ 500 eV), and those at high energy which contribute to long range damage.

Furthermore, we propose attaching a cold atomic and molecular beam source to the ARM to enable the study of medically relevant radioisotopes in the gas phase and discuss simulated results of the combined detector-source assembly. Two recently developed technologies make multi-electron coincidence detection and compatible radioactive sources practical for this approach. First, the use of a cryogenic buffer gas beam (CBGB) *[31, 32]* allows for access to unperturbed, low-energy AEs with well-defined initial conditions. The CBGB can provide a dense, cold, and well-collimated source of neutral refractory elements in the gas phase, allowing for a wide variety of elemental and even molecular species to be accessible to this apparatus. Crucially, collection of sufficient statistics from experimentally viable sample activity levels (<< 1 Ci) is enabled by the high extraction efficiencies and comparatively low beam velocities of a CBGB source. Second, new atomic layer deposited microchannel plate electron multipliers using a borosilicate glass microcapillary array allow for the simultaneous measurement of many electrons with good spatial and excellent temporal resolution. Importantly, they can be uniformly manufactured with large detection areas, increasing overall detection efficiencies, while maintaining low background rates *[33]*.

The basic layout of the ARM, shown in figure 1, starting from the CBGB source to the detector system is as follows: First, a pulsed laser ablates a solid target inside the CBGB containing the radioisotope of interest. Following ablation, the plume of atoms is entrained in an atomic beam of cryogenic helium cooled to a few Kelvins and collimated by a skimmer. After the skimmer, an electrostatic deflector deflects any ions or electrons that may be in the beam before continuing into the detection region. The detection region, inspired by the technique of cold target recoil ion momentum spectroscopy (COLTRIMS) *[34]*, is defined by two pairs of parallel-facing microchannel plate (MCP) detectors separated by a uniform electric field which is normal to the MCPs and perpendicular to the atomic beam. Ideally, the strength of the atomic source will be tuned such that on average one atom will decay in the detection region per ablation pulse. In practice, however, the only requirement will be to vaporize of order 10-100 mCi of material to detect order $10^3$ coincidence events before a significant fraction of the sample has been lost to decay. Ablation from a pulsed laser can be tuned to vaporize several milligrams of material or more per hour and, depending on the specific activity of the sample, accumulation of sufficient statistics can be accomplished using one or multiple samples. Following an Auger decay, there are up to tens of electrons emitted in much less than a nanosecond leaving the decayed species in a positively charged state equal to the number of emitted electrons. The electric field directs the electrons to one MCP detector and the ion to the other to enable coincidence detection. Finally, the atomic beam enters a shielded beam dump to reduce potential backgrounds.

We have completed construction and initial demonstration of the detection chamber and progress on the assembly of the cryogenic buffer gas beam source following the design of *[35]* is presently underway. To reflect this, we separate discussion of the completed detection chamber from the proposed integration of a radioactive beam source into different sections. In section 2 we describe the detection chamber hardware, data acquisition, and data processing capabilities. In section 3 we discuss a proof-of-principle demonstration of the ARM system to simultaneously measure electron multiplicities in logarithmically spaced energy bands of AEs following X-ray photoionization of stable krypton beyond the K-edge. In section 4 we introduce our cold buffer gas beam design for supplying radioisotopes into the ARM detector and discuss the constraints it places on charge state resolution and statistics. Finally, in section 5 we simulate and compare data analysis for a hypothetical measurement of a radioisotope with the proof-of-principle stable Kr measurements. For application of the ARM with a radioactive source, we limit our discussion to the simulation results of a beam of $^{191}$Pt passing between the detectors. However, our design is quite general and is expected to provide characterizations for a large variety of radioisotopes.

## 2. The ARM detection chamber, data acquisition, and real-time data processing

Reconstruction of the timing and location of Auger events in the detection chamber is made possible through coincidence detection of the emitted electrons and the leftover ion. The detection chamber, pictured in figure 2, comprises two pairs of parallel facing microchannel plates—the top one for detecting ions (iMCP) and the bottom one for detecting electrons (eMCP). The MCP pairs (Incom, 108 mm × 108 mm) are mounted in chevron configuration between two silver-coated ceramic (Macor) mounts used for biasing the MCP detectors. The iMCP and eMCP are separated by 12.5 cm and each MCP is paired with a metal mesh mounted at a 1 cm distance from their respective faces. The volume between the mesh and iMCP pair defines a charge acceleration region used to increase the impact energy of ions and maximize detection quantum efficiency of the iMCP. The meshes are connected in series via 4 MΩ resistors with 6 evenly spaced metal wire hoops which are used to define the outer boundaries of the detection region and ensure generation of a uniform electric field for accurate time of flight reconstruction. Independent control of the applied voltage to each mesh allows for a tunable electric field magnitude between 0 and 300 V/cm. This way, the voltage and gain of each MCP pair are independent from the applied electric field. Additionally, a tunable uniform magnetic field parallel to the electric field is generated by a pair of external coils in Helmholtz configuration to assist in electron confinement within the detection region.

The iMCP is read out via a two-dimensional delay-line anode (DL80 from RoentDek) for position and timing resolution of the ion impact (see figure 2). Signals from the delay line and the cathode of the iMCP are capacitively read out. On the electron side, the eMCP is read out via a printed circuit board with conducting pads which collect locally emitted electrons from the back of the eMCP. In the first proof-of-principle demonstration reported here, we performed readout with only a single pad covering the full extent of the eMCP. Position sensitive pads are used instead of a delay-line readout to accommodate the higher multiplicity of electrons and their comparatively short (< 50 ns) times-of-flights (TOFs) which are shorter than the delay-line time (80 ns) and instrument channel deadtime (150 ns). Preliminary experiments underway use a grid of 16 uniform conducting pads (2.5×2.5 cm$^2$) though the number of pads can be further increased for higher spatial resolution. Each readout signal is amplified (ORTEC FTA820A) and passed through a constant fraction discriminator (CFD, CAEN V812B) to ensure consistent timing before being recorded by a multi-channel time-to-digital converter (TDC, CAEN V1290A)). Coincidence detection of four delay-line signals from the iMCP triggers the TDC to store the previous 10 μs and following 1 μs of timing data as a single trigger event into a buffer. Events from the buffer are then transferred to a computer, enabling coincidence detection of electrons and the much slower ions, which arrive at the iMCP hundreds of nanoseconds to several microseconds after the electrons reach the eMCP.

Generally, a triggered event will contain multiple signals produced by the TDC in the form of a timestamp and an attached channel name. To characterize the Auger emission spectrum, it is necessary to discern signals from noise, group signals correlated to a single Auger cascade event, and identify the particle sources of each signal to complete full kinematic event reconstruction. As the ARM is designed to measure radioactive sources, it is subject to background from decays outside the beam. Robust background identification and suppression are thus critical. Furthermore, real-time data filtering and visualization is necessary for monitoring data taking and enabling rapid optimization of experimental parameters.

To address these needs, the data acquisition software includes clustering capabilities and real-time visualization features. Here, we apply the agglomerative clustering algorithm from the scikit-learn Python module *[36]* to identify matching ion and electron signals within a user-specified time threshold and group them within ion and electron clusters. The correlation between the timestamps of these clusters, alongside the measured particle multiplicity, allows for efficient classification of an event and discrimination of background. Visual feedback of data via the pyplot module allows for the fine tuning of the hardware parameters. Real-time filters can be applied to the data prior to the visualization to monitor specific features of the collected data, like the spatial and temporal distribution of the collected events.

## 3. Demonstration of the coincidence detection chamber

Proof-of-principle experiments with photoionization induced Auger cascades of krypton were performed at beamline 7-ID *[37]* of the Advanced Photon Source (APS) at Argonne National Laboratory. The goals of these experiments were to calibrate the timing and characterize the behavior of the instrument over a range of field settings, demonstrate the capability of the instrument for determining electron multiplicity, and highlight advantages of the ion-electron coincidence detection. For each experiment, we maintained room temperature Kr at a partial pressure of a few µTorr in the vacuum chamber and directed a narrow, pulsed X-ray beam from the APS through the center of the detection region, inducing photoionization of krypton along the path of the beam (cross section $0.1\times0.1$ $mm^2$) and roughly simulating the path of a radioactive cryogenic beam source. We correlated a timing signal referenced to the X-ray pulses with eMCP and iMCP signals to acquire precise (sub-ns) electron and ion TOFs. X-ray energy was tuned just beyond the krypton *K*-edge at 14.331 keV to maximize the electron multiplicity. Krypton was chosen for demonstration and calibration of the instrument for several reasons. As a noble gas, krypton is easily introduced into the detection region in the gas phase in atomic form. Furthermore, excitation near the *K*-edge has been well-characterized previously with Kr exhibiting charge states up to 12+ *[38, 39, 40]*.

### *3.1 Assigning electron energies*

Figure 3 compares the ion and electron TOFs for applied electric fields of 180, 100, and 20 V/cm at lower MCP gain settings and again with 20 V/cm with increased MCP gains. In the electron TOFs, two narrow peaks appear at short times in the 20 V/cm data followed by a broader peak which is present for all fields. There is some indication of the two early peaks in the higher fields, however the relatively low signal at shorter times and stronger overlap with the broad peak makes them more difficult to resolve. These three observed peaks are qualitatively attributed to three logarithmically spaced electron energy regimes (high: 10 to 14 keV, intermediate: 1.1 to 2 keV, and low: <300 eV) in the krypton emission spectrum. The high energy regime corresponds to Auger emissions originating from decay into *K*-shell vacancies and photoelectrons from photoionization of $n\geq2$ orbitals. Given the large energy difference between the 1*s* and 2*s* binding energies (14.3 and 1.9 keV respectively), these electrons have a relatively narrow range of energies between 10 and 14 keV. Energy conservation implies that there can be at most one electron in this regime per X-ray absorption. Intermediate energy regime electrons correspond to Auger emissions following decays into *L* shell vacancies from $n\geq3$ orbitals. Here, the 3*s* binding energy (~300 eV) is comparable to the $n=2$ energy splitting which results in another relatively narrow range of energies between 1.1 and 2 keV. Excluding the higher order possibility for $n>2$ initial *L* shell vacancies, there can be at most two intermediate electrons per absorption. Most electrons will be in the low energy regime. Low energy electrons have less than 300 eV and include near-threshold 1*s* photoelectrons, Auger emissions following decays into higher shell vacancies, and Coster-Kronig *[41]* and shake-off electrons.

To assign the electron TOF spectrum to the three energy regimes we plot the simulated TOF versus radial displacement iso-energy curves in each energy regime in figure 4 and show the relative detection probability along each curve for each field setting. Furthermore, we integrate the relative detection probabilities and normalize the result to give the collection efficiency of electrons as a function of energy in figure 5. For the high energy regime, only those electrons with initial velocities pointing toward the eMCP are within the detector's acceptance, otherwise the electrons will exit the detection volume sideways or be lost to impact with other surfaces and remain undetected. Thus, these electrons are exclusively detected at the earliest peak seen in the electron TOF plots of figure 3. The small area of the peak is also consistent with the low collection efficiency and abundance of trajectories expected for this regime. A similar explanation can be given for assigning the second peak to the intermediate electrons in the 20 V/cm field settings.

Here, the initial velocity is lower resulting in a longer TOF and greater density of trajectories in the detected phase space. Thus, a greater abundance of intermediate electrons and a higher collection efficiency explain the relative enhancement of the second peak from the first one. For the 100 and 180 V/cm electric field settings, the magnetic and electric field combination is strong enough to collect some of the intermediate electrons with a velocity component pointing initially towards iMCP side of the detector region. For these trajectories, the detected electrons end up arriving *after* the low energy electrons. In the higher fields there is also greater overlap between the intermediate and low energy electrons in their TOFs. Thus, instead of the distinct second peak seen in the low field settings, we observe a shoulder at early times and a tail at late times that corresponds to the intermediate regime. In contrast to the higher energy electrons, the arrival time of the low energy electrons is much more sensitive to the strength of the applied field. This is clear in the increased separation and broadening of the third peak relative to the earlier peaks as the electric field is reduced. At these low energies, initial velocities in all directions can reach the eMCP which contributes to its unique asymmetry and high overall collection efficiency (see figure 5). The significantly enhanced collection efficiency combined with the high abundance of the low energy electrons explains the prominence of the third peak.

### *3.2 Detector efficiency corrections*

For the ion TOFs in figure 3 there is a noticeable shift in the ion charge distribution towards lower charge states with decreasing electric field settings. Again, a shift toward lower charge states is observed as both iMCP and eMCP gains are increased for the 20 V/cm electric field setting. These shifts can generally be explained by changes in the coincidence detection probability of the ion and its emitted electrons. For the ion detection probability of the MCP, Krems et al found that a universal, monotonically increasing curve emerged when plotting quantum efficiency versus the impact energy divided by the square root of the ion mass *[42]*. Somewhat counterintuitively, the impact energy of the ions at the iMCP is *reduced* with increasing electric field settings for our detector voltage configuration. This reduction occurs because changes in the electric field are accomplished by adjusting the voltage at the ion mesh surface. As a consequence, increasing the electric field in the detector region results in a decrease of the electric field in the ion accelerator region and vice versa. The net effect is a reduced impact energy with higher fields and thus the shift in the distribution toward lower charge states with decreasing field settings is not unexpected. Furthermore, we later discovered that our CFD threshold for the low gain settings rejected a significant fraction of true signals. As the pulse height distribution of the MCP is dominated by the impact kinetic energy and not the charge state *[43]*, lower impact energy ions are less likely to have a pulse height above detection threshold even if they do generate a pulse. Thus, for the low iMCP gains used in these data, the decreased pulse heights for low charge states likely disproportionately suppresses detection of lower charge states beyond that described by Krems et al. For the electron detection probability, we would expect the opposite trend in ion TOF charge state distribution with electric field as decreasing the electric field will decrease the electron detection probability. However, we mitigate the impact of the electric field on detection efficiency with an applied magnetic field. In each case, the applied magnetic field was 20 G which limits differences in the collection efficiency of the electrons between data sets (see figure 5) and thus shifts in charge state distribution can largely be attributed to the changes in the ion detection efficiency.

When we increase the iMCP and eMCP gains at the 20 V/cm setting, the impact energy is not significantly changed compared to the low gain setting, yet the shift in the distribution to lower charge states is still clearly apparent. Further enhancement of the pulse height distribution by increased iMCP gains likely contributes to this additional shift, however, the most significant contribution can be explained by the increased detection probability of the coincident electrons. Under these settings, we observe an enhancement of the background signal attributed to $H_2^+$. For singly charged hydrogen $E/M^{0.5}$ is comparable to krypton with a charge state of 9+ which suggests the relative enhancement of its signal compared to $Kr^{9+}$ is a consequence of the changes in electron detection efficiency rather than changes in ion detection efficiencies. More evidence for the enhancement of electron detection probability can be seen in a comparison of the electron TOF plots in figure 3. Because instrument deadtime (150 ns) limits each CFD channel to only registering the first pulse from an electron above threshold, higher probabilities of detection will bias the distribution toward earlier electron arrival times. In the data with the larger gains, the two early peaks are more prominent and the median of the slower electrons is shifted earlier.

The observed charge state probabilities are shown in the hashed bars of the histogram in figure 6 for the high MCP gain and 20 V/cm field settings with applied magnetic fields of 0, 20, and 30 G. To interpret these probabilities as the electron multiplicity distribution, we must apply correction factors accounting for the iMCP quantum efficiencies as a function of impact energy and for the relative detection efficiency of detecting at least a single electron. The former means applying the universal curve discovered by Krems et al *[42]* and the latter can be approximately treated as a Bernoulli process. In this model the number of trials in the Bernoulli process is equal to the multiplicity $m$ and the probability of success for each trial is given by the single electron detection efficiency $p$ such that the probability of detecting at least one electron from an ion with charge state $z=m$ is given by $1-(1-p)^m$. We observed the 20 and 30 G data to be roughly similar to each other. This is expected as most electrons will be below 100 eV, where the single electron collection efficiency is 0.75 for 20 G and 0.9 for 30 G (see figure 5). After accounting for the open area ratio of the MCP (0.6), the detection efficiencies are $p$=0.45 and $p$=0.54. Thus, for $m$=4 the correction factor is already less than 10%. In contrast, the collection efficiency for the 0 G data is 0.2 and the correction factor is more than a factor of 2 for $m$=4. This significant difference explains the systematic reduction in probability for detection of low charge states and, consequently, the over representation of high charge states.

### *3.3 Electron multiplicity analysis*

With applied corrections accounting for the quantum efficiency dependence on the ion impact energy *[42]* and for the electron detection probability in the solid bars of figure 6 we see self-consistency between the high MCP gain, 20 V/cm data sets and good qualitative agreement with experimental results reported by Krause and Carlson *[40]* as well as Hartree-Fock calculations by Kochur et al. *[44]*. Small differences in the measured distributions, particularly at the tails of the distribution, may be attributable to the higher X-ray energy used by Krause and Carlson (17.5 keV versus 14.331 keV) which can sometimes generate a second photoelectron in addition to the 1*s* electron. Interestingly, in contrast to the broad peak at $m$=5 with a shoulder at $m$=8,9 reported by Krause and Carlson, as in the theory by Kochur et al., two distinct peaks at $m$=5 and $m$=8 are observed. Here, we highlight the capability of the ARM to investigate the structure of the multiplicity distribution in detail. Kochur et al. assert that these two peaks in the multiplicity distribution are the consequence of a bifurcation in the beginning of the Auger cascade depending on whether an initial *K*-shell vacancy decays radiatively (distribution peaks at 4 and 5) or non-radiatively (distribution peaks at 8 and 9) *[44]*. This explanation offered by Kochur et al. is corroborated in figure 7 where we observe the disappearance of the fast peak attributed to Auger electrons from *K*-shell vacancies in the electron TOF for charge states ≤ 5 and appearance of the peak for charge states ≥8.

More evidence for this source of the bifurcation appears in the width of the ion TOF distribution as a function of charge state. Given the narrow X-ray beam size (0.1 mm), the width of the ion arrival time, $\sigma_t$, is dominated by the initial velocity distribution and expected to scale like $2m\overline{v}/(q\mathcal{E})$ where $\overline{v}$ is the initial velocity distribution width, $q$ is the charge state, and $\mathcal{E}$ is the applied electric field. In figure 8, we compare the field normalized $\sigma_t$ as a function of charge state to the width expected from an exclusively thermal distribution. We find good agreement between the two until $m$=6 where the measured widths begin to exceed the expected thermal distribution. This discrepancy can be attributed to the large recoil imparted by the single 10-14 keV electron which occurs whenever there is a non-radiative decay into a *K*-shell vacancy.

## 4. Cryogenic beam source and full instrument assembly

A 3D model of the proposed full ARM assembly is shown in figure 9 (as compared to the schematic in figure 1). The cryogenic buffer gas beam source is connected to the detection chamber by a differentially pumped deflection chamber used to remove ions from the beam. In the cryo-chamber a sample, e.g., $^{191}$Pt, is irradiated by a pulsed laser that generates a plume generally containing any kind of possible products from the evaporated atoms. Pulse properties can be tuned to maximize the quantity of neutral atoms which are then entrained into the cryogenic helium beam and extracted with efficiencies exceeding 50% *[31]*. Though optimized for neutral production, charged species may still be present, which could potentially overwhelm the detectors. Therefore, we will employ an electrostatic deflector to remove any ions from the beam before it enters the detection chamber. The presence of neutral clusters produced during ablation is also a possibility. To mitigate the role of clusters we will probe the contents of the beam with a quadrupole mass spectrometer (QMS) and optimize for atom production.

In contrast to the experiments performed with krypton, the thermal contribution to the initial velocity distribution is negligible in the cryogenic beam; however, the contribution from the recoil momentum will remain and the width of the atomic beam will be substantially larger than the X-ray beam. For atomic beams, the resolution of charge states in the ion TOF data will be dominated by the fractional uncertainty of the distance of the decayed ion from the iMCP which is approximately equal to the fractional uncertainty in identifying the charge state for each ion TOF. Therefore, accurate determination of the charge for the high multiplicity Auger emitters we envision measuring require that the beam diameter must be of order 2 mm or smaller (i.e., a charge resolving power of $q/\Delta q \gtrsim 25$ ) and collimated well enough (~10 mrad) to maintain this size over the length of the detector region. To meet these stringent demands on beam quality we collimate the beam using a small-aperture (~ 1 mm) skimmer cone.

An important consideration in the design is the total sample activity required to obtain enough statistics for a measurement. Assuming a radioactive sample is completely ablated, the integrated activity passing through the detector region, $A_d$, will be related to the sample activity, $A_s$, by the net extraction efficiency of the beam, $\epsilon$, via the relation $A_d = \epsilon A_s$. The total number of decays in the detector region will therefore be $A_d t_d$ where $t_d$ is the average transit time of an atom through the detector region. Thus, the length of the MCP detector region and slow velocity of the CBGB are important design features for collecting sufficient signal from a radioactive sample. With typical velocities ~150 m/s, the average transit time through the detector region will be ~0.5 ms. Assuming total extraction efficiency after the skimmer to be of order $10^{-3}$ and requiring of order $10^3$ detected decay events, we estimate a total requisite activity of 10-100 mCi of the target isotope. This could be accomplished in a single target or spread out over several in a sequence if needed. Because coincidence detection, spatial filtering, and ablation timing suppress background, the greatest constraint on the signal rate comes from the half-life of the sample itself, i.e., we need to acquire our signal before the sample decays away. With assistance from the real-time data filtering and visualization software, we anticipate tuning the ablation parameters to achieve an average of 1 decay per pulse, however we have the flexibility to reasonably operate as low as 1 decay per $10^4$ to $10^5$ pulses running at the 20 Hz repetition rate of our ablation laser (in principle the rate would only be limited by the extraction time of the buffer gas, i.e., of order 10 ms). For medically relevant isotopes, the half-life is of order a few days meaning that anywhere from $10^2$ to $10^8$ radioactive atoms passing through the detection chamber per pulse are sufficient to meet the required signal rate. Thus, the ablation conditions are relatively modest to achieve and ultimately afford many orders of magnitude flexibility in setting a usable signal rate. Therefore, the primary constraint for collecting sufficient signal for characterization will be the total activity of the sample and not the per pulse throughput.

## 5. Simulated detector performance with a cryogenic beam of $^{191}$Pt

In section 3 we demonstrated the capacity of the ARM to accurately determine electron multiplicities of photoionization induced Auger cascades and provided an example into how the detailed structure of the multiplicity distribution can be investigated by identifying electron energies on a logarithmic scale and correlating observation of high energy electrons to particular charge states. One of the primary goals of our instrument is to characterize radioactive Auger emissions by determining their multiplicity distribution and electron energy distribution as demonstrated in the experiments performed on stable krypton. The radioactive case, however, differs from the stable experiments in two significant ways: (1) accumulating statistics from a radioactive source is significantly more challenging and (2) a radioactive decay is spontaneous and therefore not synchronized to an external clock which can be used to precisely determine electron and ion TOFs.

For the first difference we note that the detection efficiency optimization and careful calibration of timing and analysis of the instrument response performed in the stable photoionization experiments are critical to maximize and accurately analyze the low statistics (~$10^3$ events) of a future radioisotope measurement with the ARM. We expect complementary experiments using stable sources will remain important for future radioisotope measurements. Furthermore, as discussed in section 2, background suppression enabled by spatial filtering along the beam path, temporal filtering for physically allowable TOFs, and coincident detection of ions and electrons is critical to preserving the quality of low statistics data.

To accommodate the lack of a synchronized stable time reference to each Auger emission, we take a statistical approach using the median detected electron arrival time. That is, the median detected electron arrival time in each event is treated as the time reference for each event. This statistical approach is made possible by the high multiplicity of Auger decays and the capacity of the ARM to detect multiple electrons per event. Moreover, the median is robust to asymmetry and long tails which may be present in the electron TOF distribution (see, e.g., figure 3) and therefore insensitive to the precise details of the true distribution.

Physically, electrons are randomly emitted in all directions and thus an electron is equally likely to begin its trajectory pointing away from the eMCP as it is pointing toward it. Thus, for perfect collection efficiency, detected electrons would be equally distributed before and after the TOF of an electron with zero initial velocity and the peak of their distribution would be precisely at the zero-energy electron TOF. In practice, most electrons will be of the low-energy, high collection efficiency variety. Thus, the median TOF of detected electrons is the most probable TOF for a given measurement and is well-approximated by the calculated zero-energy electron TOF.

The capacity for the ARM to accurately determine ion charge states and assign detected electrons to logarithmically spaced energy bins will depend critically on the quality of the median arrival time as a stable reference. Generally, the effective stability of the median arrival time will improve with increasing number of detected electrons in an event, $n$; and, in the limit of large $n$, the distribution of median arrival times asymptotically approaches a normal distribution with mean equal to the true median electron TOF, $t_{med}$, and standard deviation equal to $\frac{1}{2\sqrt{n}f(t_{med})}$ where $f(t)$ is the distribution probability density function. Using the krypton electron TOF low gain settings data shown in figure 3 as a proxy for a typical distribution we find the approximate deviation is expected to go like $1.2/\sqrt{n}$ ns, $1.6/\sqrt{n}$ ns, and $7/\sqrt{n}$ ns for the 180 V/cm, 100 V/cm, and 20 V/cm field settings, respectively. These ranges of expected deviations suggest electron energy bin assignment accuracy will be much more sensitive to field settings than ion charge state reconstruction accuracy limited by the atomic beam size.

When considering reference time stability, stronger electric fields will yield narrower electron TOF distributions and thus more precise TOF reconstructions. Similarly, stronger magnetic fields will increase collection efficiency (see figure 5) and consequently reduce the expected deviation further. On the other hand, as shown in figure 4, weaker fields will generally provide more separation in phase space between iso-energy curves. To optimize accuracy, we need to consider the size of the eMCP readout pads and the location of the magnetic field node with respect to the logarithmic energy regimes we wish to investigate. For our first radioisotope measurement we anticipate using a grid of 16 square 2.5 cm eMCP detector pads. In figure 10 we plot logarithmically spaced iso-energy curves using 100 V/cm, 12 G field settings which complement our choice of energy regimes for the pad resolution.

For these simulations we choose three logarithmically spaced energy bins based on how they are modeled and their effects in biological media. The high energy regime (>500 eV) roughly groups those electrons which deposit their energy over long distances and contribute to non-local damage. The intermediate energy regime (30-500 eV) describes electrons which contribute to local damage (~1-30 nm) and typically constitutes the majority of emitted electrons. The low energy regime (<30 eV) represents the range of energies where cross sections are subject to quantum mechanical effects. For the low energy regime, represented by the blue curve in figure 10, the maximum displacement for the low field settings is comparable to the 2.5 cm width of the pads. Thus, low energy electrons will all remain within one pad of the decay position and arrive in a narrow time window around the median TOF. In contrast, electrons in the intermediate energy (between blue and green curves in figure 10) will tend to be displaced by less than one pad width only if they arrive significantly earlier or later than the median arrival time. Electrons in the high energy regime (between red and green curves) are distinguishable from other energies mainly by their fast arrival times.

To characterize the expected performance of our apparatus design sourced by a radioactive cryogenic beam and analyzed using the statistical time reference approach outlined above, we simulate the reconstruction of electron energies and multiplicities using 1000 randomly selected $^{191}$Pt Auger decay events following the methods of Lee et al. *[45]*. The simulated reconstruction is performed in 3 stages. In the first stage the trajectory of each emitted particle and ion is calculated for each of the 1000 events. Here, we randomly assign an initial velocity and position of the decaying atom inside the detector region assuming a diverging (10 mrad) 3 K atomic beam beginning 12 cm upstream out of a 1 mm × 3 mm elliptical skimmer (shorter dimension along the vertical axis) such that the beam dimensions inside the detector region are broadened to approximately 2.5 mm × 4.5 mm. Next, the emitted particles are assigned a random direction and the recoil on the leftover ion is computed. In the second stage, the response of each detector is simulated by Monte Carlo methods using the manufacturer-published noise characteristics, dead-time response, and MCP quantum efficiency. In the third stage, the median electron arrival times are used to approximate the TOF for each particle in each event. Finally, the reconstructed multiplicity and electron energy distribution is computed and compared to the input decay events.

In figure 11 we compare the simulated inputs for the multiplicity and detected electron counts distribution for each of the three energy regimes with the reconstructed distribution. We find there is generally good agreement between input and reconstructed distributions. With accuracy limited by the vertical atomic beam dimension, the true ion charge (i.e. total electron multiplicity) is determined correctly about 71% of the time; 28% of the time there is an error of +/-1; and the largest error observed was +/-2. For the detected electrons, we find that the energy bin is assigned with 85% accuracy. Furthermore, the reconstructed electron distribution for each energy regime closely matches the input distribution which suggests that there is no strong systematic bias present in the reconstruction.

As the pad size is reduced, there is less of a need to increase the phase space separation between iso-energy curves. Thus, for smaller pads, stronger fields can be applied to increase the median electron arrival time stability. To better assess the limits of the ARM with improved readout technology, we repeat the simulation using $4\times10^6$ pads with 0.5 mm widths using the maximum field settings (300 V/cm, 50 G). For comparison, in figure 12 (a), we show the 2D density plot of a perfectly reconstructed electron displacement versus TOF distribution correcting for beam size, temperature, resolution, and recoil effects. In figure 12 (b), we reconstruct the simulated detector data using a hypothetical stable time reference for each event (e.g. a detected coincident X-ray). As expected, the net result is a blurring of the true distribution seen in figure 12 (a). In figure 12 (c), we restrict the reconstruction to using the statistical median arrival time method and find that a bright vertical line at the mean median TOF appears as an artificial artifact from the assignment of the median arrival time to the zero-energy TOF time. Nevertheless, we find that the existence of discrete iso-energy curves in each energy regime is still preserved, albeit with additional broadening.

## 6. Conclusions

We have described a new apparatus designed to measure detailed multiplicities, i.e., the multiplicity of emitted electrons following radioactive Auger decays binned by different energy regimes. With access to the previously unmeasured low-energy electrons and multiplicities, the instrument will provide new benchmarks for atomic relaxation models of isolated atoms in the regime where calculations are most challenging and disagreement is greatest *[45, 9]*. As interpretation of data is relaxation model independent, detailed multiplicities have the potential to improve confidence and accuracy of models that contribute to oncological dosimetry estimates.

The basic detector capabilities of the ARM have been demonstrated via X-ray photoionization and work is presently underway to attach a cryogenic beam source and enable study of radioactive decays. The generality of a CBGB source enables application to a wide variety of isotopes of therapeutic interest which typically have half-lives of 1 to 20 days (e.g., $^{161}$Tb, $^{197m/g}$Hg, $^{119}$Sb, $^{67}$Ga, etc.) *[46, 1]*. Most species will be amenable to laser ablation making half-life and availability at sufficient activity levels the primary constraints to perform a measurement on a given radionuclide. Namely, the half-life needs to be of order 1 day or longer to be compatible with the cryogenic pump down time scale and the available activity level needs to be of order 10-100 mCi to acquire suitable statistics. The half-life constraint is well matched to study many potentially suitable isotopes *[47]*. In anticipation of the short-lived, low statistics considerations we have also developed robust, real-time analysis software for rapidly optimizing and tuning instrument settings during data collection.

An additional and often neglected consideration is the effect of molecular binding. The low-energy Auger electrons of interest are emitted from orbitals strongly modified by bonds with neighboring atoms. There is limited data on how bonding impacts multiplicity, but existing results suggest it is significant *[38, 48, 49, 50]* and tends to increase the multiplicity. In addition to the increased multiplicity, there could also be the creation of other energetic, reactive, and potentially therapeutic fragments. For most targeted Auger isotope medical treatments, the isotope would be attached to a carrier molecule to deliver the radioisotope within range of the cancerous cell. The effects of the carrier on the Auger emitter, however, are not well-characterized.

Originally developed as a high-density source of ultra-cold molecules, the CBGB source can be tuned to deliver certain molecules containing Auger emitters. Though we have restricted our discussion to analyzing an atomic source, the high spatial resolution of the delay-line and long ion TOF time scales lends itself well to characterizing the masses and charge states of molecular ion fragments resulting from Coulomb explosion *[48, 50]* and potentially even the total multiplicity in cases where not all ion fragments are detected by way of momentum conservation. As isotopes are typically delivered to target cells via, e.g., a chemotherapy drug, our apparatus may be able to directly explore these molecular effects and better inform theoretical models, for example by introducing a beam of cisplatin, i.e., the medically relevant form of platinum with chemical formula $PtCl_2(NH_3)_2$. Such a beam has already been demonstrated using the MALDI (Matrix Assisted laser desorption/ionization) technique *[51, 52]* and is compatible with the CBGB source.

We note that the ARM will be adaptable to improvement in readout technologies. A promising development is underway to produce a high rate, high spatial resolution MCP readout device for the TIXEL detector project at SLAC *[53]*. Comparable, albeit scaled up, readout capabilities would enable measurement of the AE energy spectrum, which according to our simulations (see figure 12 (c)) would allow for resolutions of a few eV in the lowest part of the energy spectrum. As the data is recorded at the individual event level, the spectra would enable new opportunities to explore a variety of correlations such as those between emission lines in the decay spectrum and potentially spatial emission patterns which could be investigated through repeated measurements on multiple samples to accumulate more statistics. Future versions of the instrument may also be improved with the addition of X-ray detectors which can provide independent determination of the decay event time and improve energy reconstruction, particularly for high energy electrons which typically have the shortest TOF and are therefore most sensitive to the precise time of the decay.

## Acknowledgements

This research used resources of the Advanced Photon Source, a U.S. Department of Energy (DOE) Office of Science user facility operated for the DOE Office of Science by Argonne National Laboratory under Contract No. DE-AC02-06CH11357. This work was supported by the U.S. Department of Energy Isotope Program, managed by the Office of Science for Isotope R&D and Production and by Laboratory Directed Research and Development funding from Argonne National Laboratory, provided by the Director, Office of Science, of the U.S. DOE under Contract No. DE-AC02-06CH11357. SHS and LY were supported by the U.S. Department of Energy (DOE) Office of Science, Office of Basic Energy Sciences, Chemical Sciences, Geosciences, and Biosciences Division, under contract DE-AC02-06CH11357.

## Figures

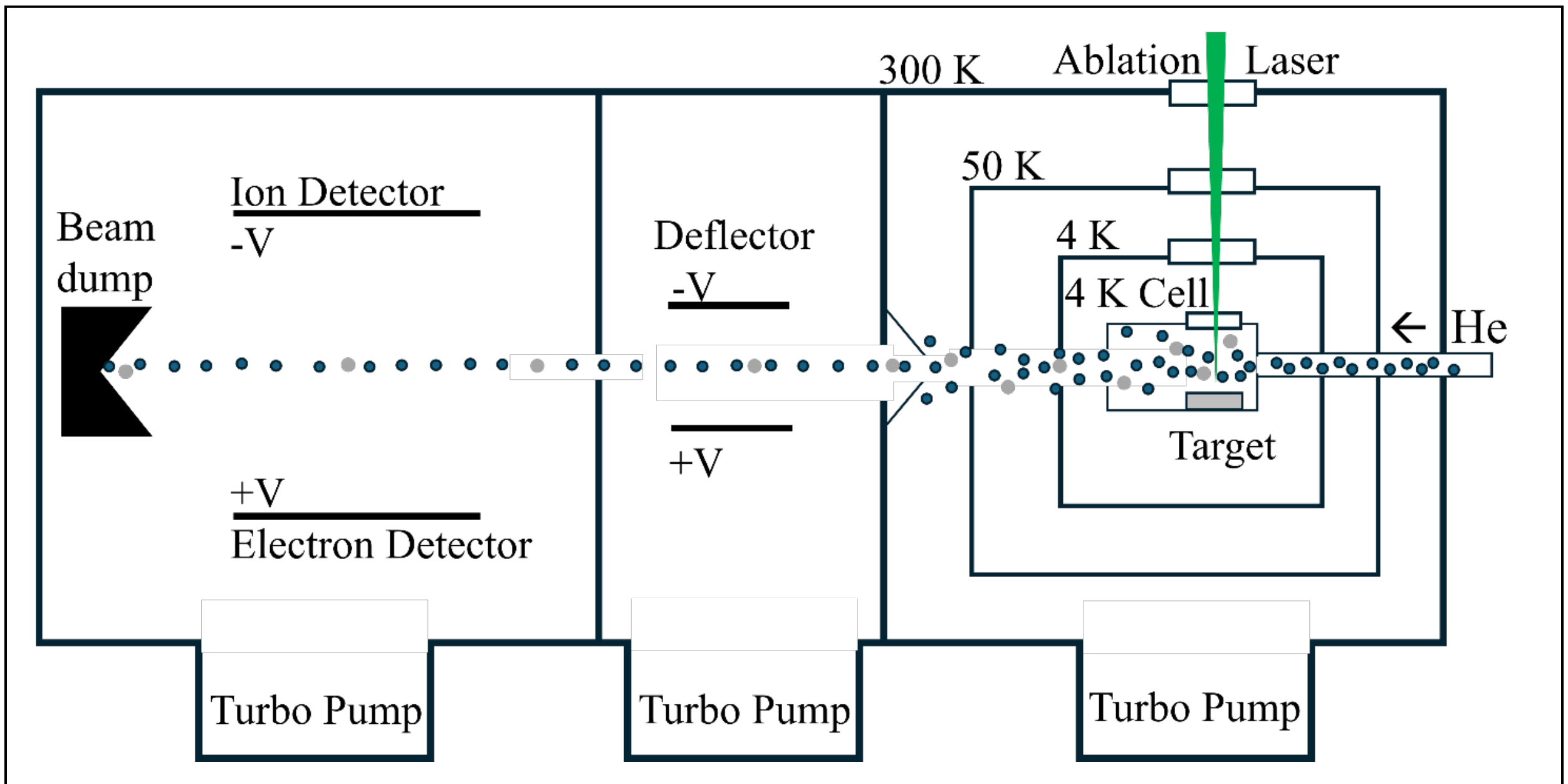

**Figure 1.** Schematic of the proposed combined cryogenic beamline and detection chamber assembly for the Auger Radioisotope Microscope. The cryogenic atomic beam (moving from right to left) containing the isotope of interest entrained in cold He is generated via laser ablation and enters the electron-ion coincidence detection chamber after passing a skimmer and ion deflector.

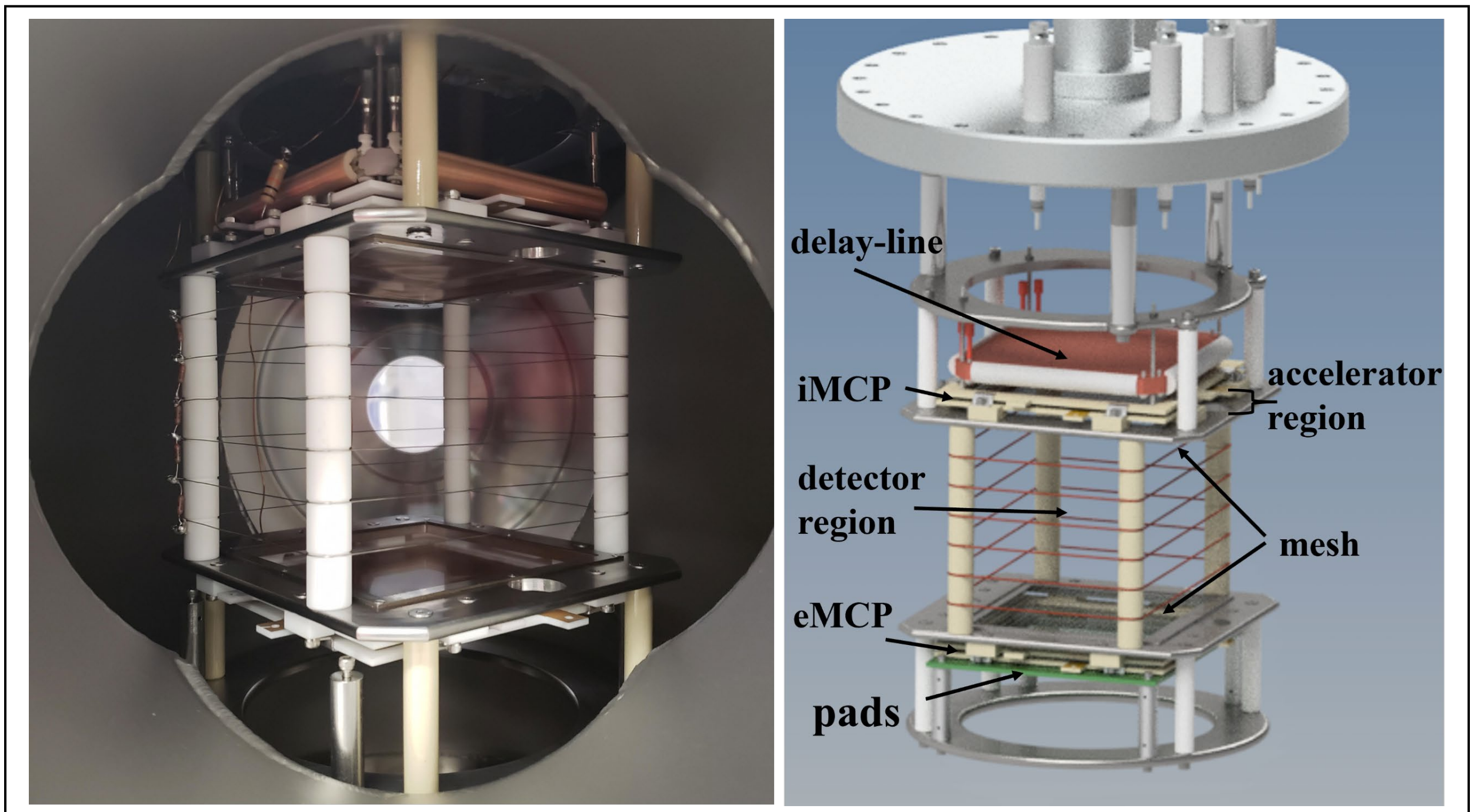

**Figure 2.** (left) Photograph of the inside of the Auger Radioisotope Microscope detection chamber with the detector region rotated for visual clarity. (right) Schematic of the detection region as defined by the mesh surfaces and the four ceramic spacers with evenly spaced steel wires. Multichannel plates on both sides of the electric field region are employed for position-sensitive ion (iMCP) and electron (eMCP) detection via delay line and charge collection pad electrode readout, respectively.

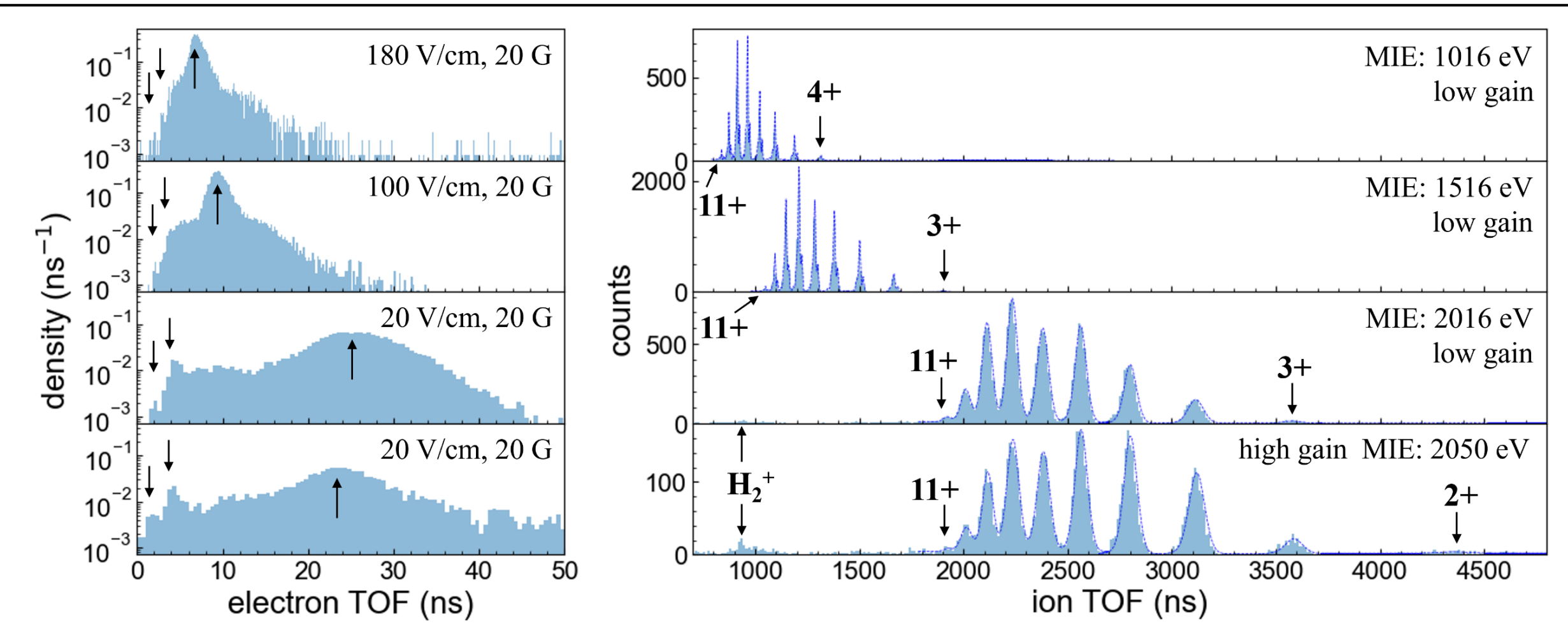

**Figure 3.** Experimental krypton coincident electron (left) and ion (right) time-of-flight (TOF) histograms following pulsed excitation from 14.331 keV X-rays. Each row corresponds to the same coincident data set. In the electron TOF column, three arrows are used to indicate peaks, from left to right, generated by three logarithmically spaced energy bands in the krypton spectrum: 10-14 keV, 1.1-2 keV, and <300 eV. In the ion TOF column, the peaks of the lowest and highest detectable charge states are identified with arrows as is a fast peak tentatively assigned to $H_2^+$ in the 20 V/cm data sets. The thin dark blue curve is a fit to the ion TOF spectrum assuming natural isotopic abundance and is used to determine peak widths and areas for figures 8 and 6 respectively. The top three rows were recorded using "low" gain settings of 1800 V eMCP and 1880 V iMCP. The bottom row used the "high" gain settings of 1850 V eMCP and 1920 V iMCP. For each unique gain and field setting combination we label the calculated minimum impact energy (MIE), i.e. the impact energy into the iMCP from a zero-energy, singly charged ion created at the center of the detector region.

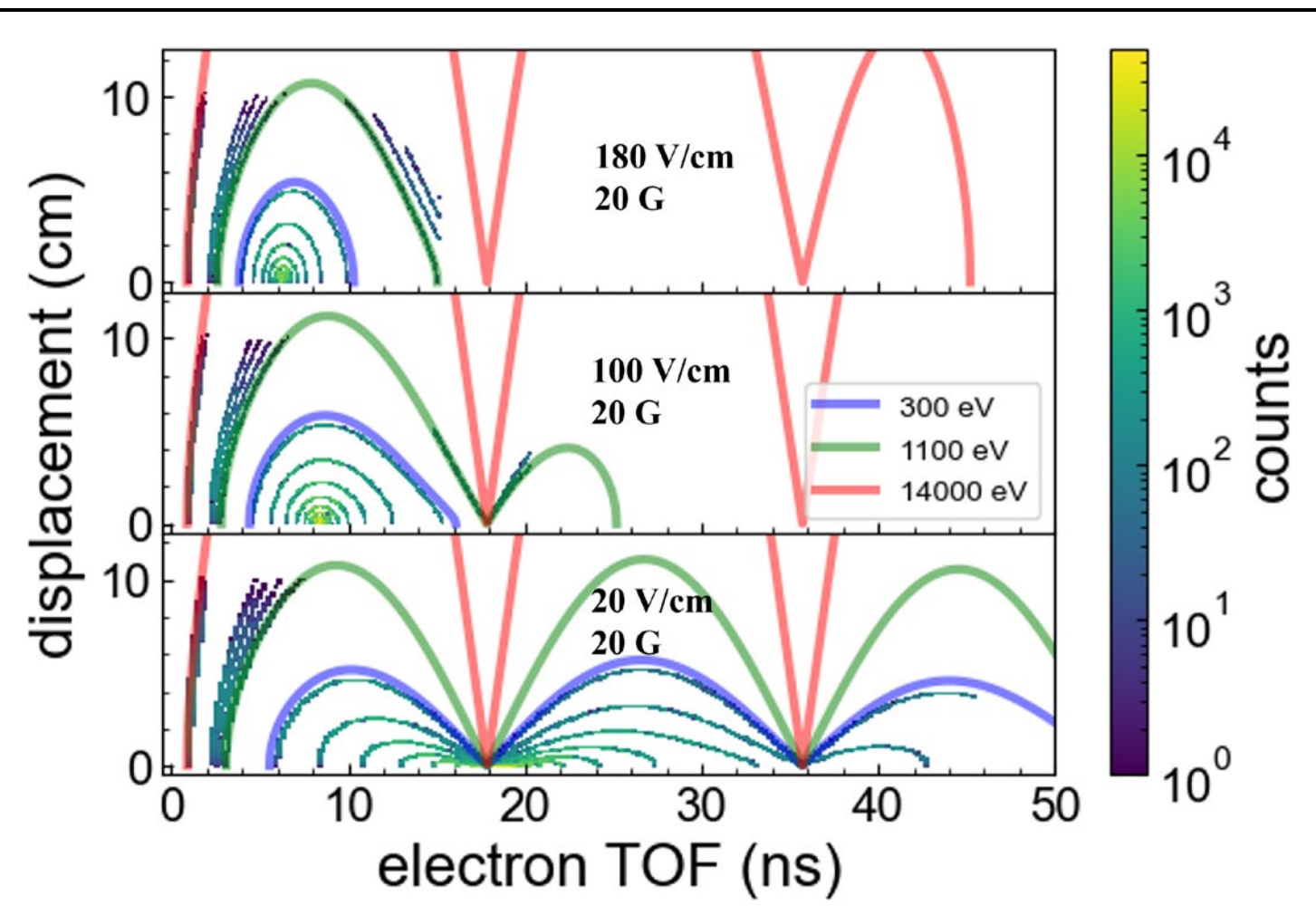

**Figure 4.** 2D density plots of electron time-of-flight versus radial displacement from initial ion position for iso-energy curves in the three logarithmically spaced bands of krypton described in section 3.1. Solid lines follow the iso-energy curves at the boundaries of the three energy regimes of krypton: 300 eV (blue), 1100 eV (green), and 14 keV (red). Simulated distributions are calculated assuming an applied magnetic field of 20 G and electric fields of 180 V/cm, 100 V/cm, and 20 V/cm (top to bottom). Each iso-energy curve density plot is generated by randomizing the initial trajectory of $10^5$ electrons along the X-ray beam axis and recording the TOF and displacement values of trajectories that successfully reach the eMCP. The color gradient along the curves, as indicated by the scale bar on the right, provides the number of counts detected for a given TOF, displacement trajectory. As displacement approaches the 10 cm length scale of the eMCP, the number of counts decreases rapidly until none are detectable. A maximum TOF is also observed as electrons with sufficient vertical velocity will hit the iMCP before reversing direction back toward the eMCP. Nodes in the iso-energy curves occur when the time of flight is equal to integer periods of the cyclotron resonance time.

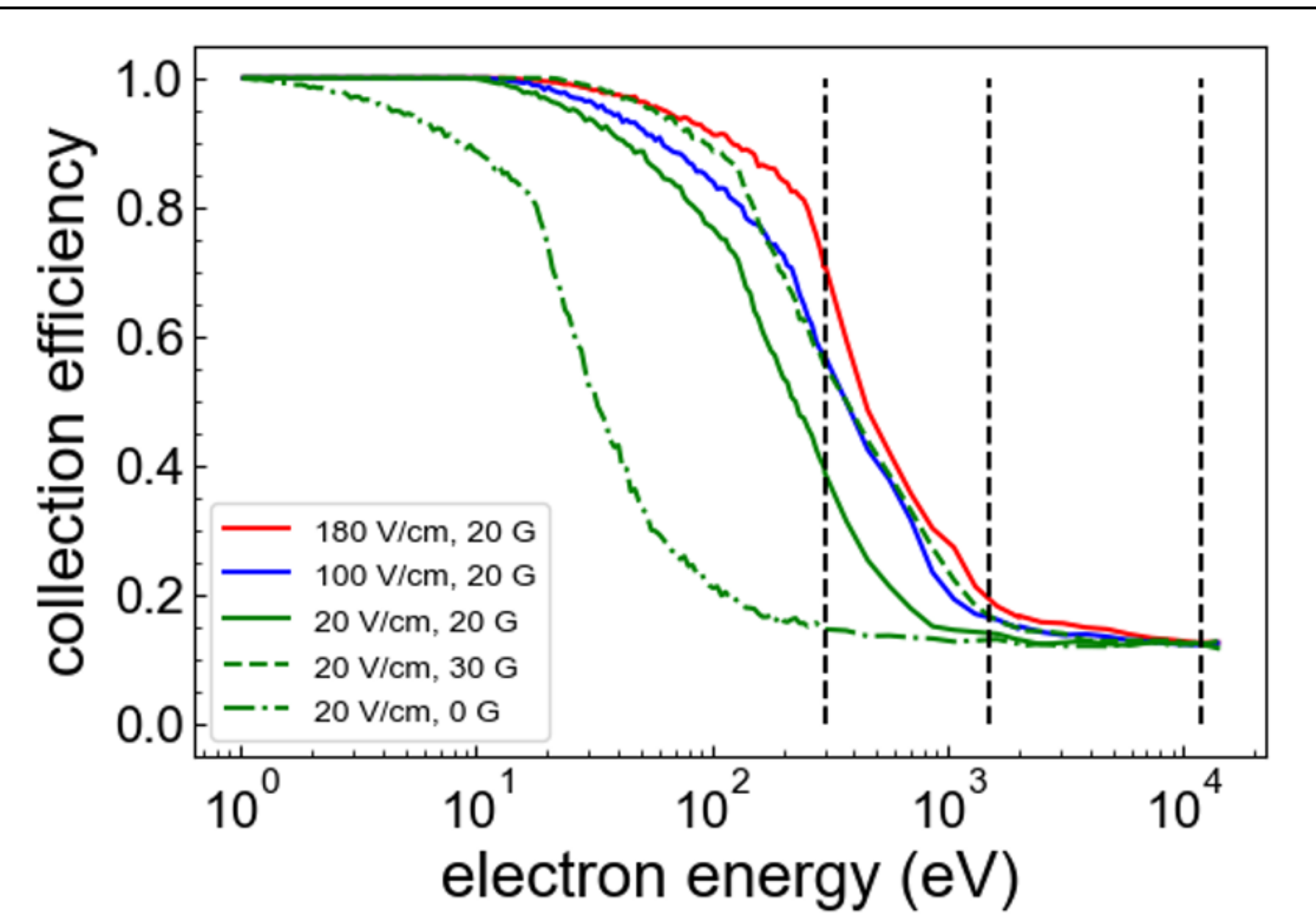

**Figure 5.** Simulated collection efficiency curves as a function of electron energy for different experimental conditions. Curves are computed using a Monte Carlo simulation to determine the fraction of electrons that reach the eMCP as a function of energy. Vertical lines at 300 eV, 1500 eV, and 12 keV are shown to guide the eye to the location of the 3 logarithmically spaced energy regimes in Kr. For the purposes of applying a correction to the multiplicity plot shown in figure 6, collection efficiencies of 0.2, 0.75, and 0.9 were used for the 20 V/cm 0, 20, and 30 G field settings. Assuming saturated detection with the eMCP, the collection efficiency is multiplied by the open area ratio (0.6) of the MCP to provide an estimate of the total detection efficiency.

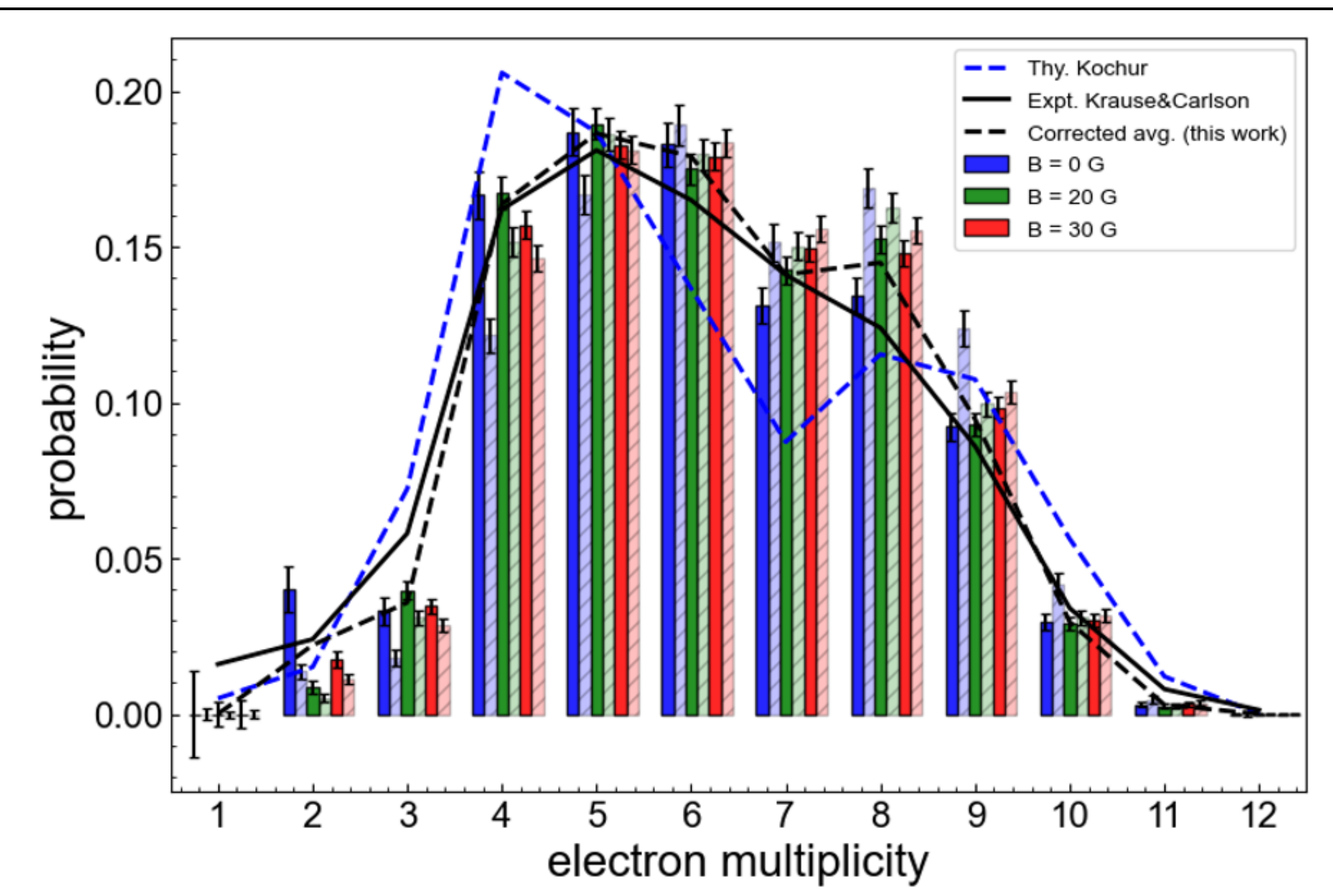

**Figure 6.** Krypton charge spectrum produced by beyond *K*-edge photoionization. Lighter, hashed bars represent the experimental normalized integrated totals for each charge state for different applied magnetic fields. Otherwise, each data set is recorded using the 20 V/cm, high MCP gain settings with X-ray energy tuned to 14.331 keV. Solid bars correct each data set for systematic effects (see section 3.2) and the average of the 3 corrected data sets are shown in the dashed black line. Theory from Kochur and experiment from Krause and Carlson were performed with X-ray energy tuned to 17.5 keV which will have some impact on the distribution.

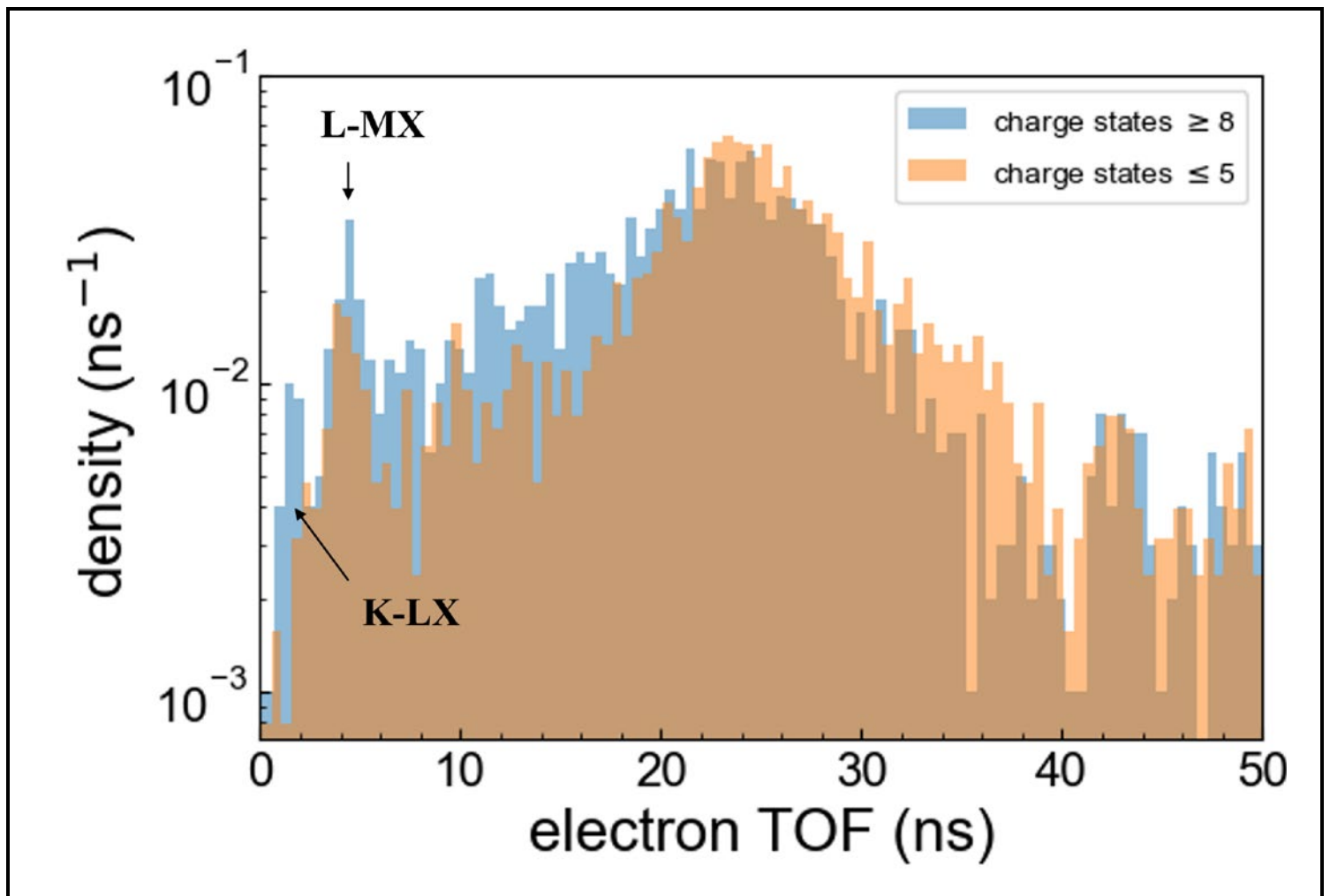

**Figure 7.** Experimental normalized electron TOF distributions for events with high (≥ 8) and low (≤ 5) ion charge states. Data acquired with high MCP gain and 20 V/cm, 20 G field settings. K-LX and L-MX bands are identified as sources for two early peaks strongly correlated with ion charge state.

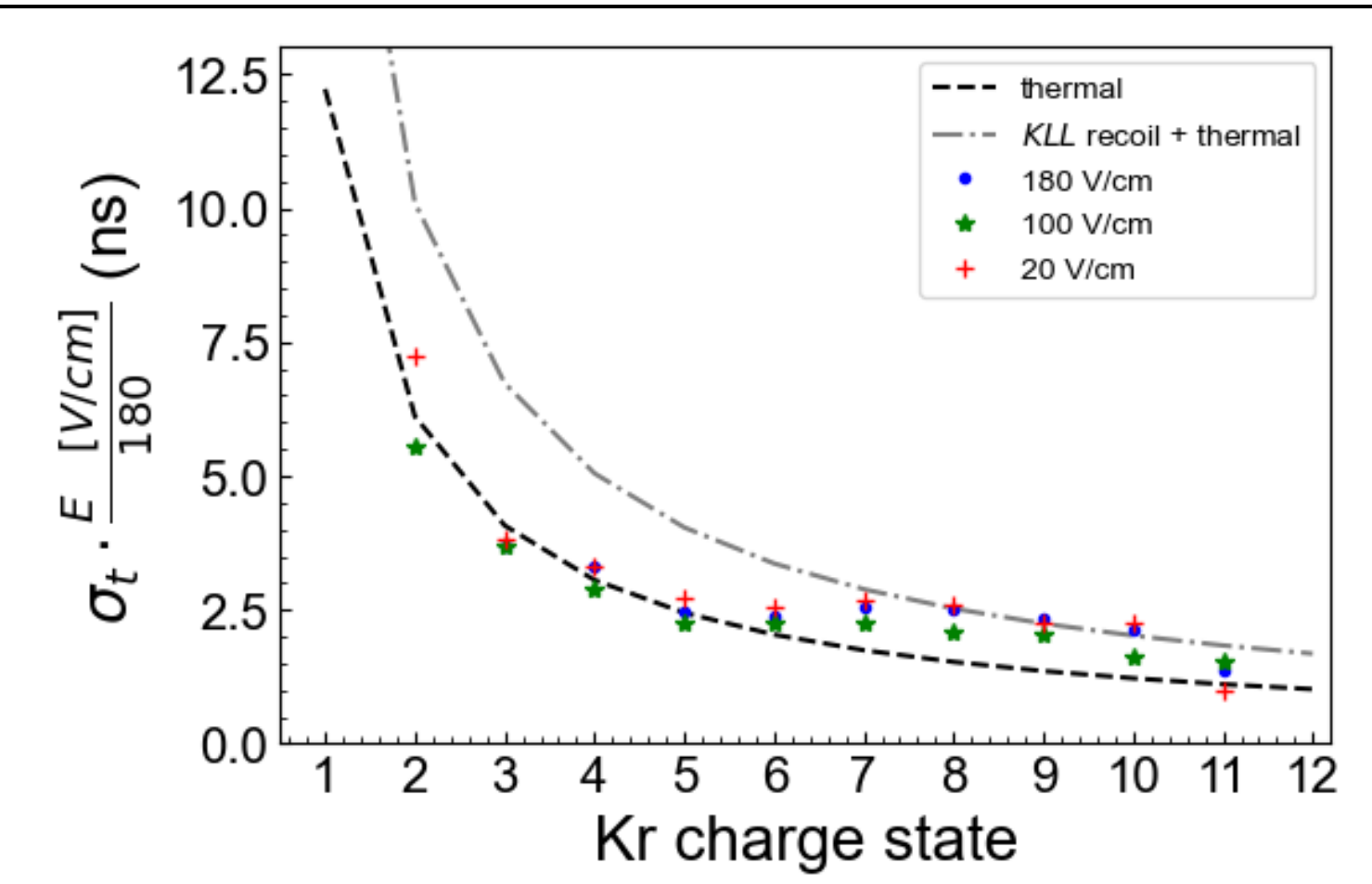

**Figure 8.** Observed charge state peak widths normalized by applied electric field plotted together versus computed *ab initio* room temperature thermal width with and without *K-LL* Auger electron recoil (i.e., recoil from an 11 keV electron). Widths diverge from the thermal-only width beginning at $q$=6 and converge to the width calculated for thermal plus recoil for higher charge states.

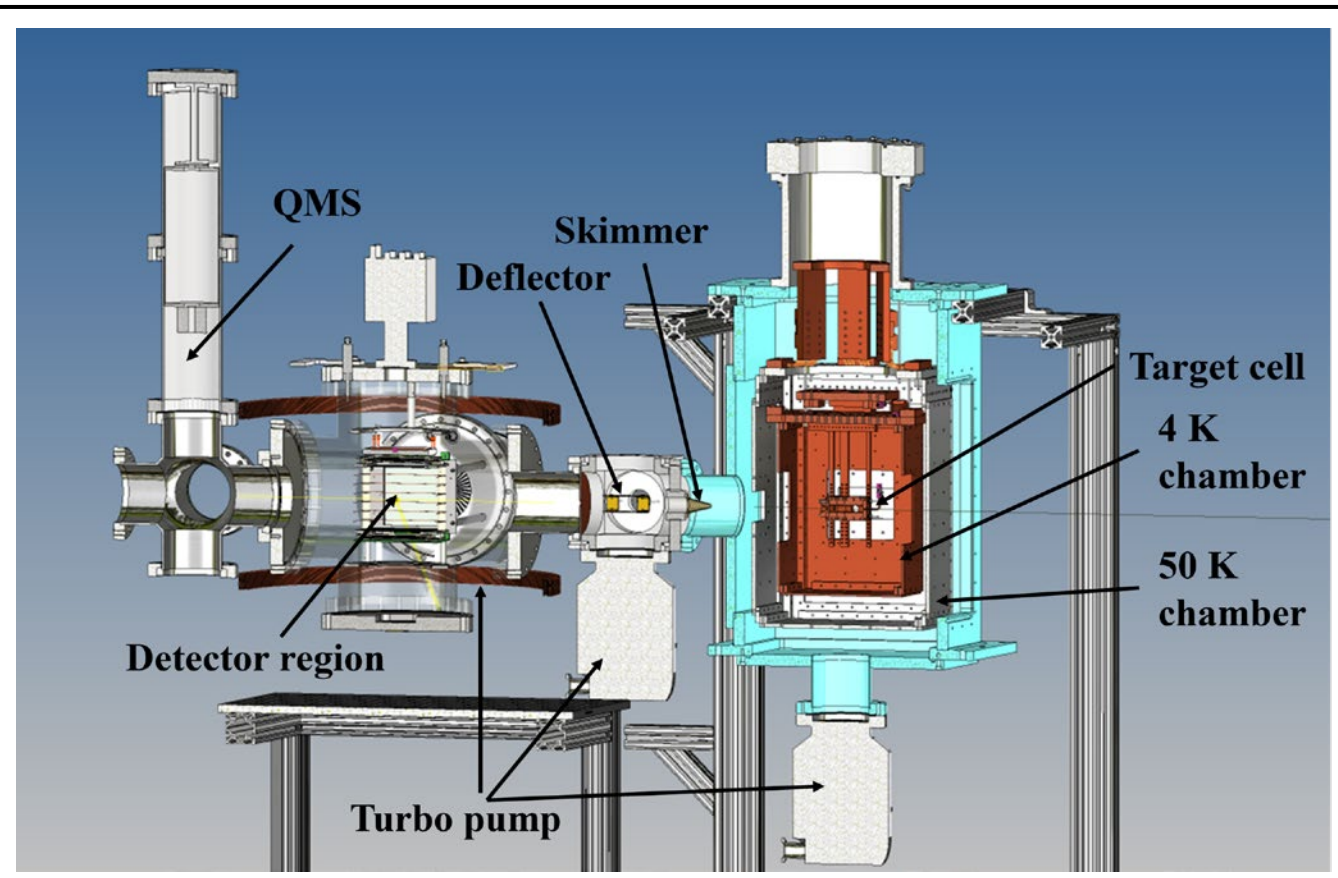

**Figure 9.** 3D model of the proposed combined cryogenic beamline and detection chamber assembly. A third turbomolecular vacuum pump is positioned behind the detector region. A quadrupole mass spectrometer (QMS) is added at the end of the beamline for additional beam diagnostics.

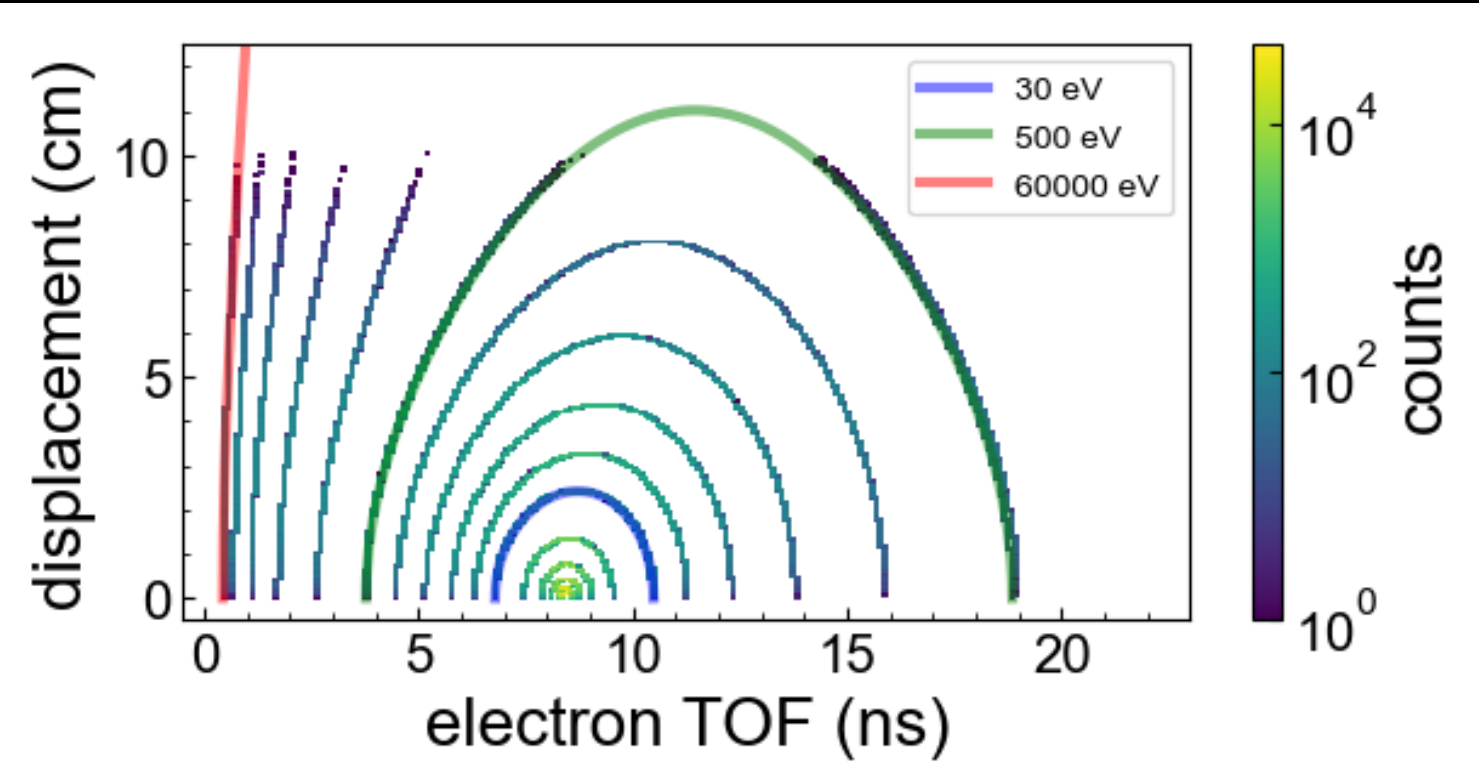

**Figure 10.** 2D density plots of electron time-of-flight versus radial displacement from initial ion position for logarithmically spaced iso-energy curves. The simulated distributions are calculated assuming an applied electric field of 100 V/cm and magnetic field of 12 G. The three solid iso-energy curves are plotted to visually distinguish the three logarithmically spaced energy bins used in electron energy reconstruction (see section 5). Field settings are tuned to optimize accuracy of electron energy bin assignment by matching the maximum displacement of the low energy electrons (< 30 eV) with the size of electron detection pads (2.5 cm). As in the plots of figure 4, for each iso-energy density curve, $10^5$ randomly initialized trajectories are computed and the final displacement and time-of-flight of the trajectories which reach the eMCP are recorded.

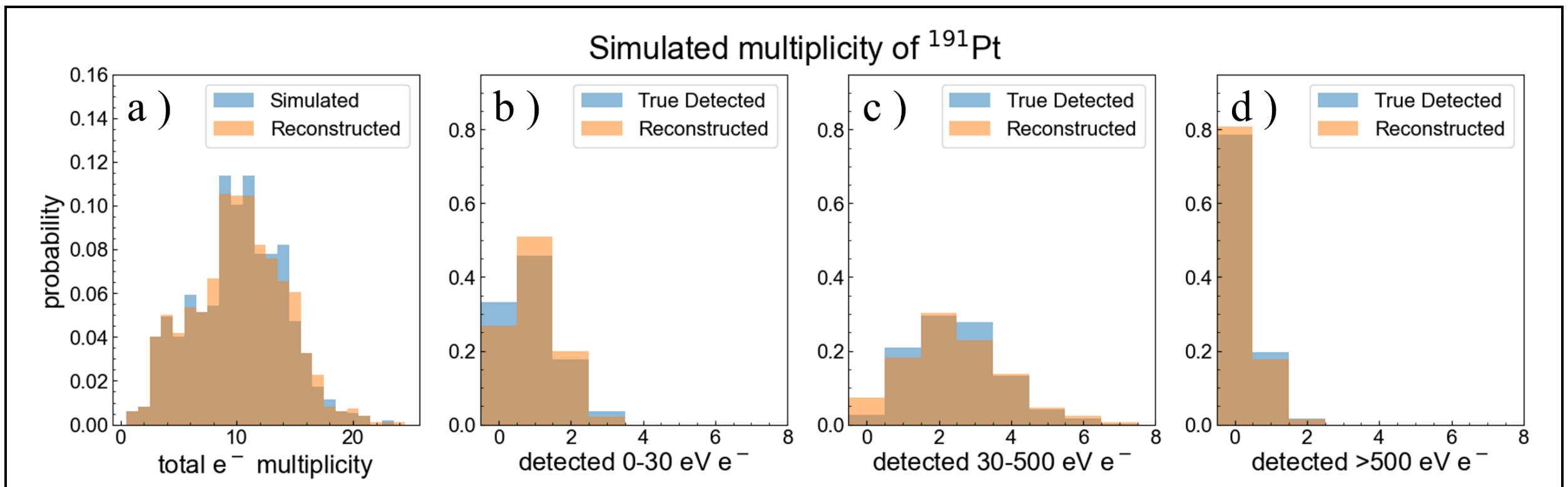

**Figure 11.** Above we compare the input distribution with the reconstructed distributions for the simulation of $^{191}$Pt decay multiplicities using a 4×4 grid of 2.5 cm square pads for readout of the eMCP. (a) Total multiplicity for 1000 simulated $^{191}$Pt decay events. The remaining plots show the distribution for number of emitted electrons detected per decay in the low (b), intermediate (c), and high (d) energy regimes. Simulations were performed with applied fields of 100 V/cm and 12 G. For these reconstructions we used the median electron arrival time method described in section 5.

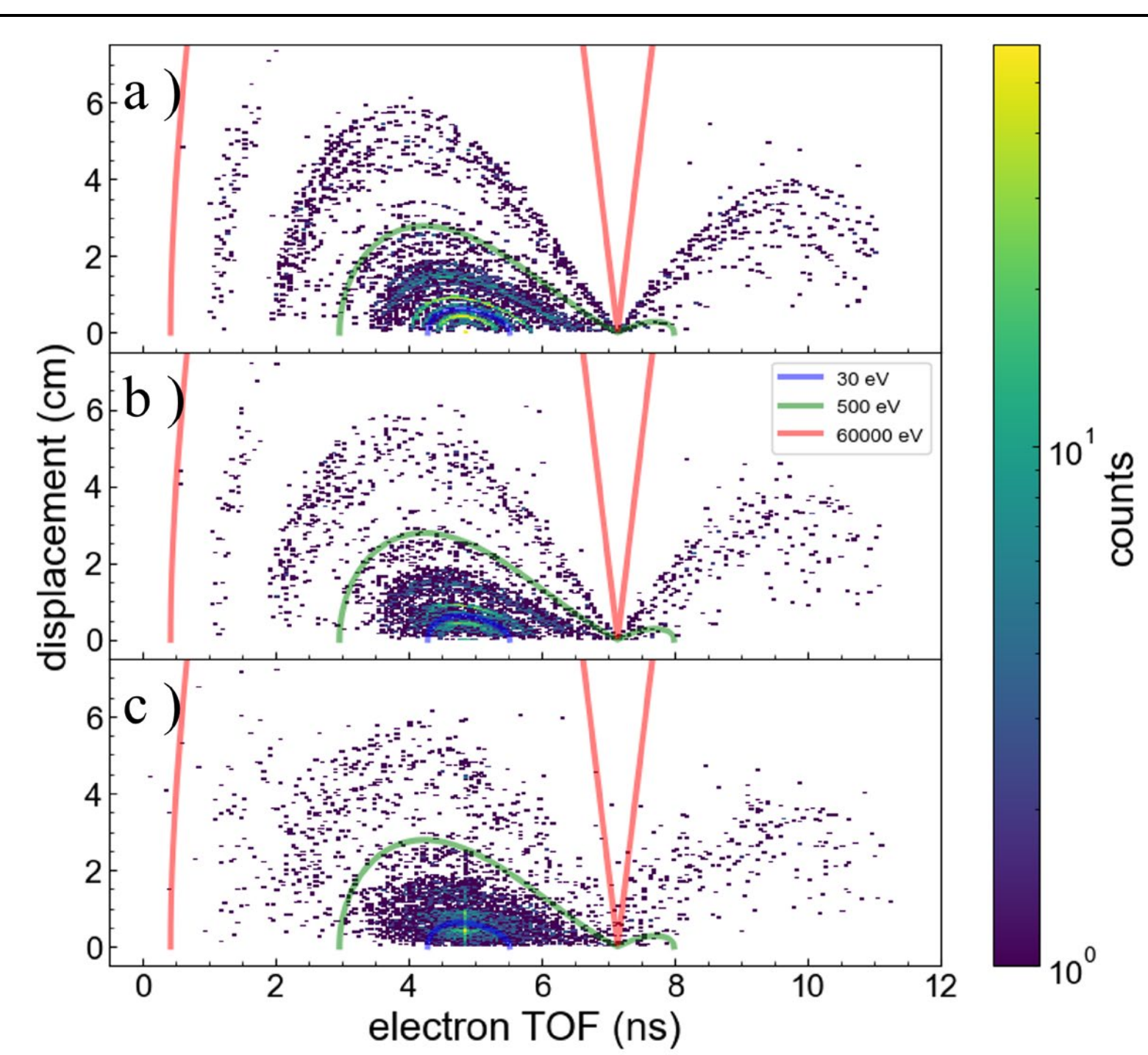

**Figure 12.** Simulated electron TOF versus displacement density plot for 1000 $^{191}$Pt decay events using applied fields of 300 V/cm and 50 G. In (a) we compute analytic trajectories assuming electrons are emitted randomly along a line. In (b) we run the full cryogenic beam source simulation and reconstruct the density plot assuming a 2000×2000 grid of 0.5 mm square pads for readout of the eMCP using a hypothetical stable time reference for each event (e.g. a detected coincident X-ray). In (c) we reconstruct the same simulated data in (b) using the median arrival time method (see section 5).